# Bending of Spaghetti Beams and Columns Exposed to Hot Steam Reveals Several Physical Properties of the Spaghetti

Fathan Akbar and Mikrajuddin Abdullah[(*)]

*Department of Physics, Bandung Institute of Technology,*

*Jalan Ganesa 10 Bandung 40132, Indonesia*

[(*)]Email: mikrajuddin@gmail.com

Abstract

In this paper, we show that bending of spaghetti beams and columns exposed to hot steam reveals the time evolution of young's modulus, the diffusion coefficient of water molecules penetrating the spaghetti, the partial Fickian behavior of water diffusion, and a logistic-like evolution of column bending angle. The bending geometries were timely recorded and the Young's moduli were obtained by processing the images. We applied two equations proposed by us previously, one equation was applied for beam bending and the other for column bending, to estimate the Young's moduli. Experiment were conducted by exposing the freely-hung cantilever spaghetti beams and columns using hot steam from a boiling water so that the images were recorded realtime while the beam or column bent undisturbedly. Surprisingly, the estimated diffusion coefficient of water molecules mathced well the experimental data reported by others. This method may become an altenative for estimating the diffusion coefficient of vapor moleculs penetrating the materials.

# I. INTRODUCTION

Deformation of spaghetti and other pastas is still being a challenge for many researchers [1-7]. This exploration opens unlimited trials and errors for structural design in a very easy way, with low risk and low cost. Using spaghetti, we can "create" materials having wide Young's modulus window by controlling the water content penetrating inside. Several authors have investigated the time evolution of spaghetti's Young's modulus and have observed many interesting and unexpected behaviors [1,3-7], the finding of which may lead to scientific and technological impacts.

Research on the evolution of pasta and noodle mechanical properties have generally been conducted by immersing such materials into a boiling water and the measurement was performed by taking them out from the water [1,8,9]. The problem faced by this approach is when one measurement has been applied to one spaghetti rod, it may destroy the rod and such a rod cannot be used for further measurement. This problem can be solved by wetting the spaghetti rods with a steam from boiling water, instead of directly immersing them in the boling water. This strategy permits the wettening of the free spaghetti rod without disturbing it and the bending images are contaclessly taken realtime. The elastic modulus is estimated by processing the images while the rod progressively bends undisturbedly. This method is also capable of wetting the spaghetti with a gas flow or wettening with hazardous liquids by generating steam from the liquids. The liquid and heater are located far away and the steam is guided with a hose to the measurement location. In addition, the images of the bending rods can also be recorded using a remote camera for savety purposes.

The purpose of this work is to investigate the time evolution of Young's modulus of spaghetti wetted by steam from boiling water. Both cantilever beams and columns are investigated. The Young's moduli are obtained by processing the image of beams or columns being bent. We applied the model reported previously for estimating the elastic modulus of various sheets [10,11].

## II. METHOD

### A. Basic Equations

The basic equation we will use to describe the bending of homogeneous slender beams (horizontally aligned) and columns (vertically aligned) is a sequential equation [10,11]

$$\theta_j = \theta_{j-1} - \frac{a^3 \lambda g}{YI} \sum_{i=1}^{j-1} i \cos\theta_i \qquad (1)$$

with $g$ is the acceleration of gravitation, $Y$ is the Young's modulus, and $I$ is the area moment. The rod of length $L$ is divided into $N$ identical segments, $a = L/N$. The mass per unit length, $\lambda$, is assumed to be uniform. $\theta_j$ is the angle made by the $j$-th segment with respect to the horizontal direction, $j = 1$ is the index for the free end segment, and $j = N$ is the index for the clamped end segment.

The boundary conditions for spaghetti beams and columns are $\theta_N = 0$ and $\theta_N = 3\pi/2$ (directs downward), respectively. Equation (1) has been successfully applied to investigate the temperature dependence of the Young's modulus and to estimate the glass transition temperature of various polymers [11].

For vertically aligned cantilever sheets we have obtained a general scaling relationship [10]

$$|\Delta\theta| \propto (1/L_0 - 1/L)^\alpha \qquad (2)$$

with $|\Delta\theta|$ is the angle made by the free end of the cantilever column with respect to the vertical line ($|\Delta\theta| = 0$ when the column is straight upward), α ≈ ½ and $L_c = (7.8373 YI/\rho A g)^{1/3}$ is the critical length for a self-bucklingcolumn. We will apply this equation also for estimating the spaghetti's Young's modulus. Therefore, two equations are used together and their results will be compared.

### B. Experimental

We used a spaghetti with a commercial name LaFonte-10 produced by Indofood Sukses Makmur, Indonesia, containing *durum wheat semolina* and *gluten*. In the package, one rod has a length 0.26 m, a diameter of 0.0013 m, and a mass of 0.0006 g. Therefore, the linear density is $\lambda$ = 0.0023 kg/m and the area moment for the cylindrical rod is $I = \pi r^4/4 = 1.4 \times 10^{-13}$ m$^4$.

We investigated the rod of lengths of 0.08 m, 0.10 m, and 0.12 m. The arrangement of measurement is illustrated in Fig. 1. The spaghetti was exposed to boiling water steam generated from a cooking pot, heated using a gas stove. We placed the spaghetti close to the water surface to minimize the deviation of the steam density with time. We assume the steam concentrations on the rod surface stayed constant. The recorded temperature of steam around the spaghetti varied in the range between 46 ºC to 50 ºC.

The spaghetti bending was recorded using a camera. Chillo et al reported the optimum cooking time of spaghetti to obtain better cooking quality for spaghetti immersed in boiling water is around 10 min [12]. However, at present work we used longer time since the spaghetti were exposed to water steam.

### C. Procedure for Estimating the Young's Modulus

For a specific image of beam bending, we manually determine the angle made by the free end ($\theta_1$). The bending angles of other segments were calculated using Eq. (1) until obtaining the bending angle of the fixed end, $\theta_N$. We search for the Young's modulus so that $|\theta_N - \theta_{fix}| \leq \varepsilon$. In this work we used ε = 0.001 rad. The procedures were repeated for other images that have been photographed at different times. For spaghetti columns, we manually measured the angle of the top (free) end relative to the vertical direction for estimating the Young's modulus. Description of the procedures and program are provided in Supplementary 1.

### III. RESULTS AND DISCUSSION

First we inspect how much the spaghetti's dimension changed by exposing it to the steam. We hung vertically the spaghetti rods so that the bottom end was closely above the water surface and exposing for 35 minutes. We observed the change in the length (100% ×Δ$L$/$L$) was less than 5% (see Supplementary 2). This value was much smaller when compared to the one reported by Del Nobile and Massera [13], the change of which was nearly 15% for spaghetti immersed in boiling water. Based on this fact we ignored the change of length and assumed the length is constant in simulations. We did not measure the change on radial dimension, but we assumed the change fraction was also nearly the same.

Figure 2 shows the estimated Young's moduli obtained from a beam of 0.12 meters long (see Supplementary 3 for the comparison of the photographed images and the simulation results). We plot the Young's moduli both in linear and logarithmic scales, while the time remains in the linear scale. The data have been obtained by searching for the $Y$ so that the simulation results using Eq. (1) best fit the image, including $\theta_N < \varepsilon$. Fitting processes were performed for each image. In calculation we used $g$ = 9.77 m/s$^2$, the data in Bandung city, Indonesia [14]. In simulation, we divided the rod into $N$ = 100 segments.

To confirm the obtained data, we compared the simulation results and the experimental data for beam of length 0.1 m. The beams of length 0.10 m and 0.12 meters have been exposed simultaneously so that they must have the same Young's modulus at the same time. The beam of length 0.08 m was exposed at the next time shift. We obtained the consistency of the fitting and theobserved result for beam of length 0.10 m when using the Young's modulus as used in simulating the beam of 0.12 m length. It indicates that the estimated elastic moduli shown in Fig. 2 are confident.

From both data we concluded that the Young's modulus decays exponentially with time up to 17 minutes. Fitting the lograrithmic data with a straight line, resulting a slope of $m = -0.133 \text{ min}^{-1}$. For a very long time, the Young's modulus must be finite, equal to the Young's modulus of the wet spaghetti. Therefore, we propose an approximated equation for the dependence of the Young's modulus on exposure times as

$$Y(t) = Y_w + (Y_d - Y_w)e^{-mt} \qquad (3)$$

with $Y_w$ and $Y_d$ are the Young's moduli of the wet and dry spaghetti, and $Y_w$ must be too small compared to $Y_d$. Other authors have also observed the same trend, but they did not account for residue ($Y_w$) [15]. Del Nobile et al measured the crystal level of spaghetti at different hydration time saturated at a finite value (did not approach zero), and this level can be assumed to be contributed by $Y_w$ [16]. The presence of the residue at long time is indicated in figures reported by Caferi et al [3].

If we exptrapolate the fitting line to $t = 0$, we shoud have the elastic modulus of the dry spaghetti as $Y_d \approx 0.67$ GPa. Several authors have reported that the Young's modulus of the dry spaghetti varies between 0.05 GPa to 5 GPa [18-20]. To demonstrate the true Young's modulus of the dry spaghetti we used here, we repeated the method conducted by Vargas-Calderón et al when measuring the Young's modulus of noodles and bucatini [6] (see Supplementary 4). The displacement of the free end satisfies $\Delta y = WL^3/3YI$, with $W$ is the load. We fit the measured data of displacement against load to obtain the slope (= $L^3/3YI$). We obtained a value of $Y_d$ = 2.25 GPa. Vargas-Calderón et al reported the elastic modulus of the bucatini of 2.33 GPa and of the noodels of 2.96 GPa [6]. Therefore, our extrapolation data for the steamed-spaghetti was smaller than the result of direct measurement. It is, therefore, means that the estimation using Eq. (3) shold not be applied at the initial time of exposure.

Let us compare the experiment result with theoretical prediction based on water diffusion into the spaghetti rod. Assuming the diffusion coefficient, $D$, is constant and water concentration depends only on the radius variable. The diffusion of water into spaghetti can be described by the Fick's law [16] and in the cylindrical coordinatesit can be expressed as $\partial C/\partial t = (D/s)\partial/\partial s(s\partial C/\partial s)$, with $C$ is the water molecule concentration inside the spaghetti rod and $s$ is the distance from the cylinder axis. The initial condition is $C(s,0) = 0$ for $0 \leq s \leq r$ and the boundary condition is $C(r,t) = C_a$, with $r$ is the rod radius. The solution, $C(s,t)$, can be expressed as [7]

$$1 - \frac{C(s,t)}{C_a} = \frac{2}{r}\sum_{n=1}^{\infty}\frac{J_0(\alpha_n s)}{\alpha_n J_1(\alpha_n r)}\exp(-n^2 Fo) \qquad (4)$$

with $J_0(x)$ and $J_1(x)$ are the Bessel function of the 0$^{th}$ and 1$^{st}$ orders, $\alpha_n$ is the root of the $J_0(\alpha r) = 0$ and $Fo = D\pi^2 t/r^2$ is the Fourier number.

Reported by Horigane et al, the diffusion coefficient of the spaghetti during boiling is 4.8 – 4.9 × 10$^{-10}$ m$^2$/sec andduring holding is 2 - 3 × 10$^{-11}$ m$^2$/sec [7]. In our results, the spaghetti started to become soft at around 10 min (600 s). Therefore, the estimated Fourier number at the time when softening starts was $Fo \approx 7$. The ratio of the second and the first terms in Eq. (4) is $\exp(-28)/\exp(-7) \approx 7.6\times10^{-10}$, which is very small. Therefore, for explaining the softening of the spaghetti rod, it is enough to use only the first term in Eq. (4) so that

$$1 - \frac{C(s,t)}{C_a} \approx \frac{2J_0(\alpha_1 s)}{\alpha_1 r J_1(\alpha_1 r)} \exp(-Fo) \qquad (5)$$

At a certain time, the concentration of water depends on the distance from the axis. The highest concentration is located on the surface and decreases when moving toward the center. Higher water concentration leads to lower Young's modulus. The dependence of the Young's modulus on $C$ can be expanded in Taylor series as $Y(C) = \sum_{n=0}^{\infty} a_n (C/C_a)^n$ where $a_n$ are constants (series coefficients). By taking only the first two terms, we will have $Y(C) = a_0 + a_1(C/C_a)$ which can be rewritten as $Y(C) = \xi(Y_d - Y_w)(1 - C/C_a) + Y_w$ with $\xi$ is the "normalization" constant. Del Nobile et. al [16] and Cuq et al [17] have proposed the dependence of Young's modulus on water content in spaghetti as $Y(C) \propto \exp(-\nu C)$ with $\nu$ is a parameter. If $C$ is small enough, this expression can be approximated as $Y(C) \propto (1 - \nu C)$, which is similar to out proposal after adding the lowest boundary (residue) $Y_w$. It is clear from this approximation that at $t = 0$, $C = 0$ so that $Y(C) = Y_d$ and at $t \rightarrow \infty$, $C = C_a$ so that $Y(C) \rightarrow Y_w$. Therefore, the time dependence of the Young's effective elastic modulus $Y(t) = \int Y(r) dA / \int dA$ becomes

$$Y(t) \approx Y_w + \xi(Y_d - Y_w) \int_0^r \left[ \frac{2J_0(\alpha_1 s)}{\alpha_1 r J_1(\alpha_1 r)} \exp(-Fo) \right] \frac{2\pi s ds}{\pi r^2}$$

$$= Y_w + (Y_d - Y_w) \exp\left[-(D\pi^2/r^2)t\right] \qquad (5)$$

after selecting the "normalization" constant to satisfy $\xi \int_0^r [2J_0(\alpha_1 s)/\alpha_1 r J_1(\alpha_1 r)](2\pi s ds/\pi r^2) = 1$. This time behavior in Eq. (5) is the same as that obtained from experiment as displayed in Fig.2 and Eq. (3).

However, we need to compare the exponential factor in experimental fitting and that given by Eq. (5). By comparing Eq. (3) and (5) we have $D = mr^2/\pi^2$. Using $m = 0.133$ min$^{-1}$ from fitting results and $r = 6.5 \times 10^{-4}$ m we obtain $D = 9.5 \times 10^{-11}$ m$^2$/s. Surprisingly, this estimation is very close to the data reported by others. Horigane et al have reported the diffusion coefficient of water in the spaghetti are $4.8 – 4.9 \times 10^{-10}$ m$^2$/sec during immersed in cooling water and $2 - 3 \times 10^{-11}$ m$^2$/sec during holding [7]. Since we have exposed the spaghetti in the water steam, the condition should lay between the two conditions reported by Horigane et al and closer to the holding condition. Doulie et al have reported the diffusion coefficient for pastais $0.8 - 9.3 \times 10^{-11}$ m$^2$/s [21]. We can then state here that, the present model can be used for estimating the diffusion coefficient of water inside the spaghetti.

Equation (2) is similar to that proposed by Chillo et al [9]. Baiano et al fit the time dependence of the Young's modulus using a simple decay fuction $Y(t) \propto \exp(-\gamma t)$, with γ is the decay constant [15].

The question might be raised concerning the accuracy of the fitting process. We demonstrated the prediction deviation of around 5% (see Supplementary 5). This deviation can be claimed to be very accurate since in general, the measurement of the modulus of elasticity gives rise to inaccurate results. The deviation up to 30% might happen. For example, the elastic modulus of human femoral trabecular bone was reported to be $11.4 \pm 5.6$ GPa (variation by 49%) using nano-indentation method [22], $0.44 \pm 0.27$ GPa (variation by 61%) using compression method [23] and $0.29 \pm 0.18$ GPa (variation by 61%) using torsion method [24]. For bone, the variation of measurement can be up to 20% [25]. By vibrating a cantilever using an electromagnetic method, Joshi et al reported a measurement of wire or strip can produce an error up to 35% [26]. In case of the dry spaghetti, the elastic moduli were reported to vary between 0.05 - 5 GPa [18-20].

For further prove of the accuracy, let us start from Eq. (1). By cosidering the first and the second segments only we have $\theta_1 = \theta_2 - (\gamma/Y)\cos\theta_1$ with $\gamma = a^3 \lambda g / I$. The vertical size of the first segment is $y_1 = a\sin\theta_1 = a\sqrt{1-(\delta\theta_{12}/\gamma)^2 Y^2}$ with $\delta\theta_{12} = \theta_1 - \theta_2$. If the Young's modulus changes by $\Delta Y$, the vertical size of the first segment changes by $\delta y_1 = d/dY[a\sqrt{1-(\delta\theta_{12}/\gamma)^2 Y^2}]\Delta Y = -aY(\delta\theta_{12}/\gamma)^2 \Delta Y / \sqrt{1-(\delta\theta_{12}/\gamma)^2 Y^2}$
$= -a(\cos^2\theta_1 / \sin\theta_1)\Delta Y/Y$. The total deviation of the free end when the Young's modulus changes by $\Delta Y$ becomes approximately $\Delta y = N\delta y = -Na(\cos^2\theta_1 / \sin\theta_1)\Delta Y/Y$ $= -L(\cos^2\theta_1 / \sin\theta_1)\Delta Y/Y$. The maximum error of estimation will be obtained when $\Delta y \approx d$ so that

$$\frac{\Delta Y}{Y} \approx -\left(\frac{d}{L}\right)\frac{\sin\theta_1}{\cos^2\theta_1} \qquad (6)$$

For example, using $d = 1.5 \times 10^{-3}$ m, $L = 0.1$ m dan $\theta \approx 30°$ we have $\Delta Y/Y \approx 2\%$.

For further confirmation of the previous results, let us analyze the bending of spaghetti colums. Equation (2) can be rewritten as $L/L_c - 1 = \gamma L |\Delta \theta|^2$, with $\gamma$ is a constant. When the spaghetti column is very soft, $Y = Y_w$, the bending angle becomes maximum, $|\Delta \theta| \to \pi$ and $L/L_c(Y_w) \gg 1$ so that $L/L_c(Y_w) \approx \gamma L \pi^2$ from which we obtain $\gamma = 1/\pi^2 L_c(Y_w)$. We can also write $L_c = \kappa^{1/3} Y^{1/3}$ with $\kappa = 7.8373 I/\rho A g$ and by defining $\tilde{L} = L/\kappa^{1/3}$, Eq. (2) can finally be expressed as

$$Y = \left[ \frac{\tilde{L}}{1 + (\tilde{L}/Y_w^{1/3})\Delta\tilde{\theta}^2} \right]^3 \qquad (7)$$

with $\Delta\tilde{\theta} = \Delta\theta/\pi$.

Figure 3 is the plot of $\ln(Y/\tilde{L}^3)$ against time for spaghetti of lengths 0.12 m, 0.10 m, and 0.08 m. In calculation we used $Y_w = 0.05$ GPa based on trend of the exponential decay in Fig. 2 that likely approximate at such Young's modulus.

All curves show a sudden decrease at certain times: $t_{c1} = 24.8$ min, $t_{c2} = 29.4$ min, and $t_{c3} = 34.6$ min for columns of length 0.12 m, 0.10 m, and 0.08 m, respectively. Those times are when the column starts to bend due to self-buckling [10]. At $t < t_{ci}$, the Young's modulus remains large so that the column height remains less than the critical height for self-buckling. As time increases, the Young's modulus decreases and the critical height for self-buckling occurs at $t_{ci}$. At that time, the Young's modulus has been low enough due to high water penetration so that the critical height for self-buckling is precisely the same as the applied column height.

For the three lengths evaluated here, we obtain that having surpassed $t_{ci}$, the data drop linearly with time (in logarithmic scale). At this region we can approximate $Y(t) = Y_w + \Delta Y \exp(-mt)$. Fitting the data in the drops region with a straight line we have $m = 0.201$ min$^{-1}$, 0.138 min$^{-1}$, and 0.423 min$^{-1}$ for columns of lengths 0.08 m, 0.10 m, and 0.12 m, respectively. Surprisingly, those values are very close to that obtained in Fig 2 of 0.133 min$^{-1}$. This concludes that both procedures are suitable for estimating the Young's modulus of spaghetti slender rods exposed to steam of boiling water.

We have to note that, the Fickian behavior for data obatained from the spaghetti beam is satisfied when the exposure time has passed a few minutes. The exponential decays with a factor of $m = 0.133$ min$^{-1}$was not accurately obeyed at the initial time of exposure. Similarly, for the spaghetti column bending, the Fickian behavior likely occured after reaching $t_{ci}$ and obeyed at the region where $Y$ changed abruptly. In the entire times, the behavior was not Fickian. There are also reports on the non-Fickian behavior of the water diffusion into spaghetti rods [7,27-29].

Let us furtherly explore the data of spaghetti column bending. Figure 4 is the plot of $|\Delta\tilde{\theta}|^2$ against time for column of lengths 0.12 m, 0.10 m, and 0.08 m. The time axis have been right shifted by $\Delta t = t_{c3} - t_{c1}$, $\Delta t = t_{c3} - t_{c2}$, and $\Delta t = t_{c3} - t_{c3} = 0$ for columns of lengths 0.12 m, 0,10 m, and 0.08 m, respectively. Such shifts make all the data coincide. By careful inspection to the data we conclude that all the data satisfy a single equation, to mean the bending evolution for all column lengths are identical, except for the starting time of bending. The equation for all length of column is general, i.e. $|\Delta\tilde{\theta}|^2(t-\tau_i(L))$ with $\tau_i(L)$ merely depends on the column length. It is then interesting to search for such equation.

From Fig. 4, the maximum $|\Delta\tilde{\theta}|^2$ is unity so that it must approach the unity after surpassing the fast stepping. The most appropriate function to represent such behavior is a logistic function. The application of the logistic function to explain the change of spaghetti behavior is not new. Del Nobile and Massera have fit the change of the spaghetti length with respect to the initial length against the boiling time with a logistic function [28]. Cafieri et al have approximated the change of the Young's modulus of spaghetti with a logistic function against the boiling time [3]. Goldberg and O'Reilly used a logistic function to explain the change of spaghetti curvature with the boiling time [1].

We use the Fermi-Dirac-like function to explain the change of $\Delta\tilde{\theta}^2$ against time as

$$\Delta\tilde{\theta}^2 = \frac{1}{1+e^{-(t-\tau)/T}} \qquad (8)$$

with $\tau$ and $T$ are paremeters that must comply the experimental data. The parameter $\tau$ is the time when $\Delta\tilde{\theta}^2 = 1/2$. This value occurs at $t \approx 37.6$ min so that $\tau \approx 37.6$ min. The slope at $\tau$ is $1/4T^2$ [30]. The approximated slope of dotted line in Fig. 4 is 0.172 min.$^{-1}$. Therefore, we obtain the approximated $T = 1.21$ min. Curve in Fig. 4 has been obtained from Eq. (7) using such parameters. We see a strong consitency between the theroretical curve with the measured data.

From the discussion above we conclude that, in Fickian region, the equation for beam bending expressed in Eq. (1) and column buckling expressed in Eq. (2) gave rise to the same conclusion that the elastic modulus of the spaghetti beam exposed to boiling water steam changes with time according to $Y(t) = Y_w + (Y_d - Y_w)\exp(-mt)$. The same equation have also been reported by other authors that had investigated the cooking of spaghetti in boiling water [12,21]. Cooking in boiling water is restricted for observing the time dependent of the properties change without disturbing the sample (pulling out from the water). In our experiment, however, the rods or beams are free to bend in any directions. Our results proved that Eqs. (1) and (2) which are initially used to explain bending of sheets are universal for any slender beams or columns.

We also demonstrated that the present method is able to accurately predict the diffusion coefficient of water penetrating the spaghetti. Therefore, this method might become an alternative choice for estimating the diffusion coefficients of molecules of wetting liquid or gas inside a bendable slender rods, either they are positioned as cantilever beams or columns.

## IV. CONCLUSION

We have succeeded to estimate the time evolution of the Young's modulus of spaghetti arranged in both cantilever beams and columns. The moduli were extracted by confronting the calculation results using the two equations proposed by us previously with the experimental data. We demonstrated that two equations had convergent to the same conclusios: predicted the same timely evolution of the elastic modulus, predicted nearly the same diffusion coefficient of water into the spaghetti rods, both of which matched very well the experimental data reported by others, and predicted the Ficking behavior of water diffusion in a certain time window. In the entire time, the deviation angle of column bending satisfied the logictic equation (the Fermi-Dirac-like function). The timely dependent of column bending angles is universal after appropriate shifting the time reference, the amount of which depends on the column height.

**FIGURES**

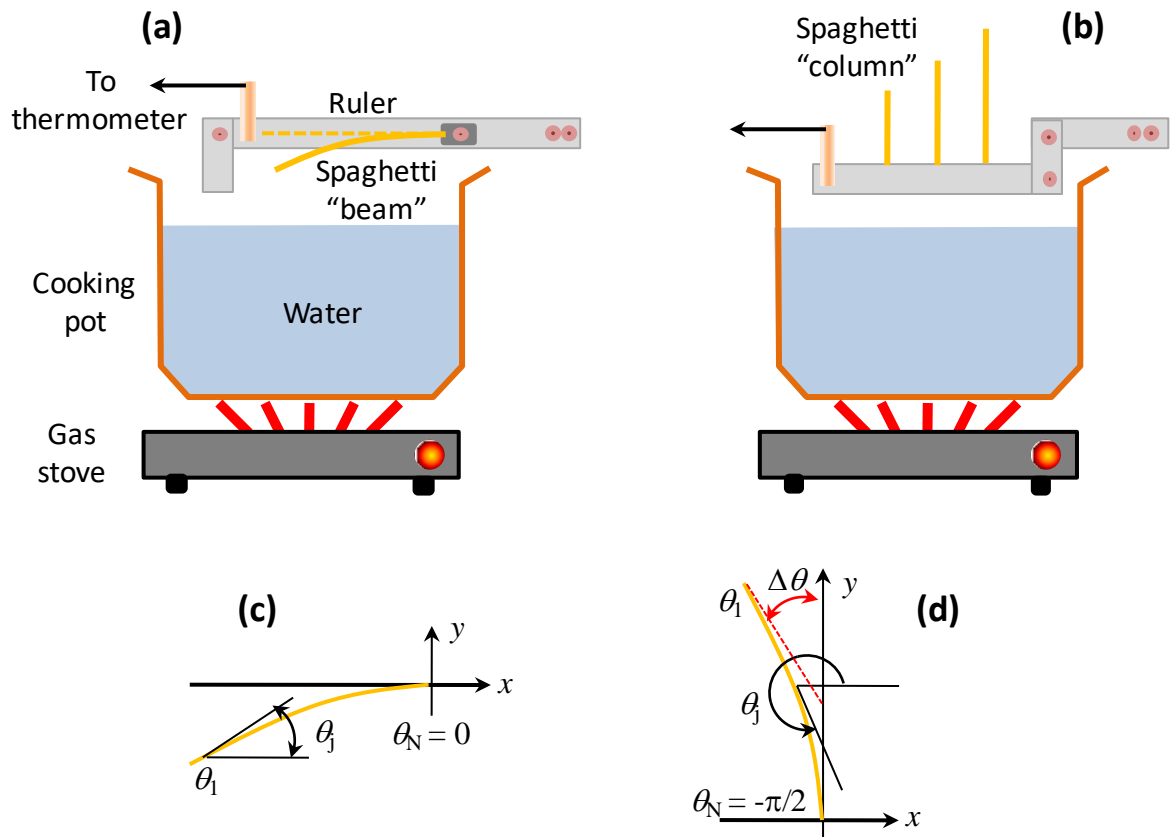

FIG. 1. Experiment setup: (left) beam bending and (right) column bending. Insets are bending profiles and the corresponding parameters for beam (left) and column (right).

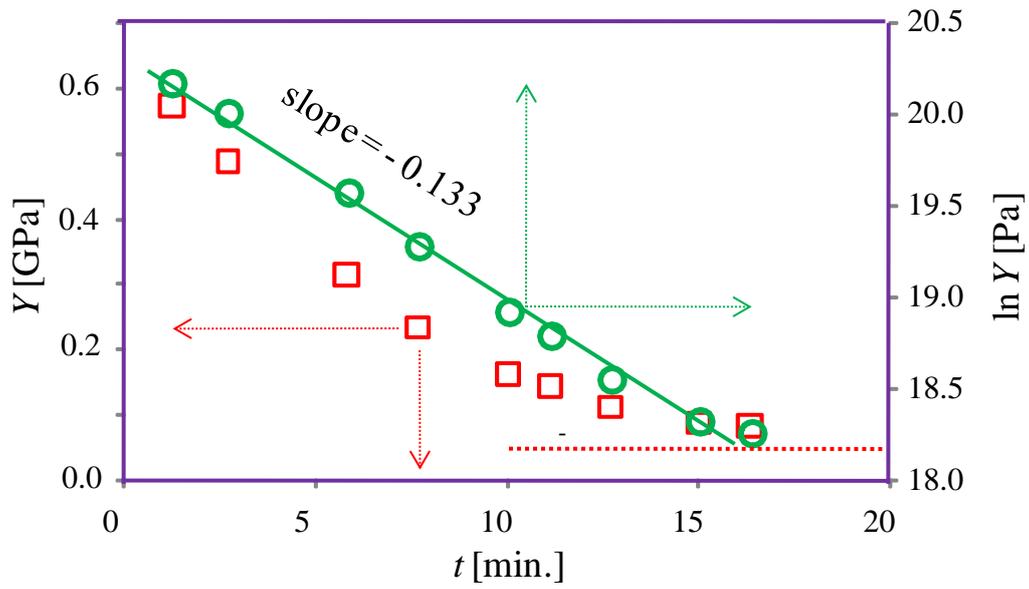

FIG. 2. Time dependent of elastic modulus obtained from the spaghettibeam of 0.12 m long. We show the data of the Young's modulus in linear scale (square symbols) and in logarithmic scale (circle symbols).

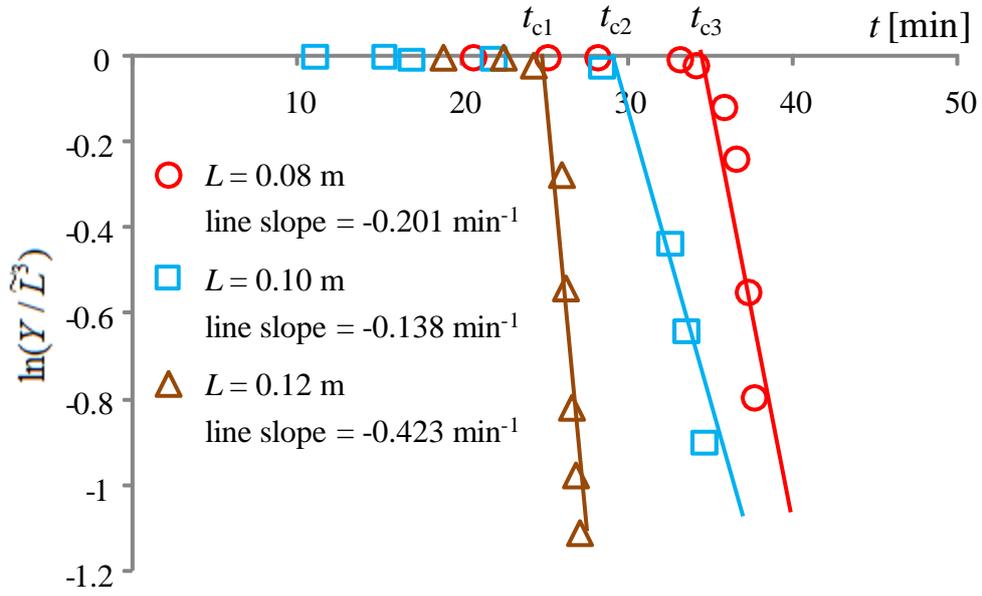

FIG. 3. The Young's modulus of spaghetti columns at different lengths: (circles) 0.08 m, (squares) 0.10 m, and (triangles) 0.12 m. The Young's moduli were calculated using Eq. (6). The $t_{ci}$ is the time when the column starts to bend due to self-buckling.

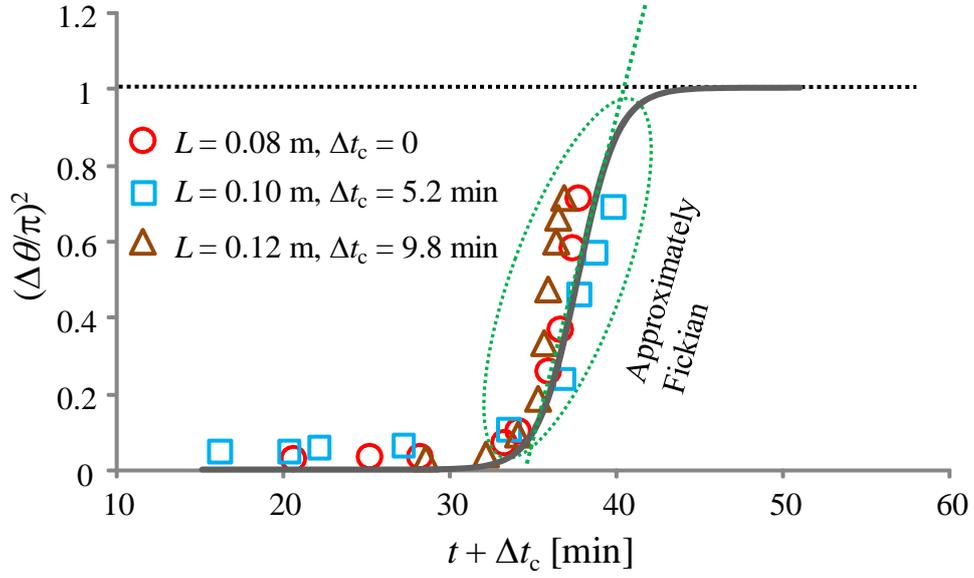

FIG. 4. The dependence of $\left|\Delta\tilde{\theta}\right|^2$ on exposure times for columns of different lengths: (circles) 0.08 m, (squares) 0.10 m, and (triangles) 0.12 m. The curve was calculated using Eq. (7) with $\tau \approx 37.6$ min and $T = 1.21$ min. The step region satisfies the Fick second law.

# SUPPLEMENTARIES

## Supplementary 1

For simulation purposes we input:

a) The angle made by the free end ($\theta_1$) that has been measured from the image
b) Other rod parameters: length density, rod length, rod diameter, accelaration of gravitaion, number of segments and the tolerable error to stop the iteration.

The process is started by inputing the first estimation of the elastic modulus, $Y_{init}$.

The iteration is stopped when the difference in the calculated angle for the fixed and and the true angle is less that the tolareble error, $\varepsilon$

The accelerate the simulation, we performed the following steps:

1) First we simulate for the first estimated young modulus and obtain $err = \theta_N - \theta_{fix}$.
2) If $err > 0$ (the estimated young modulus is larger that the true modulus) we define the following parameters

    $Y_{bottom} = 0$ and $Y_{top} = Y_{init}$.

3) Otherwise, if $err < 0$ (the estimated young modulus is smaller that the true modulus) we define the following parameters

    $Y_{bottom} = Y_{init}$ and $Y_{top} = 5*Y_{init}$.

4) At the next simulation we use the next estimation for the young modulus

    $$Y_{new} = \frac{Y_{bottom} + Y_{top}}{2}$$

5) After obtain all angles, we check again $err = \theta_N - \theta_{fix}$.
5) If $err > 0$ (the estimated young modulus is larger that the true modulus) we change the parameter $Y_{top} = Y_{new}$ and otherwise, if $err > 0$, we change $Y_{bottom} = Y_{new}$.

6) Repeat steap 4) until reaching $|err| \leq \varepsilon$

The silmutions were performed using the Visual Basic.

Visual Basic Program

---------------

```
Sub EstimatingYoungModulus()
'
Dim teta(1000), rD, rLd, d, rL, salah, areaMom, seg As Double
Dim Yinit, Ybootom, Ytop, Ynew, konst, tetaF, beda, deviasi As Double
Dim N As Integer

  bilpi = 3.141592654
  rD = Cells(3, 2) ' Rod diameter
  rLd = Cells(4, 2) ' Rod length density
  g = Cells(5, 2) 'Gravitation acceleration
  N = Cells(6, 2) ' Rod length
  rL = Cells(7, 2) ' Number of segments
  teta(1) = Cells(8, 2) ' Angle of the free end
  tetaFix = Cells(9, 2) ' Angle of the fixed end
  salah = Cells(10, 2) ' Error for itehartion to stop
  Yinit = Cells(11, 2) ' Initial estimation for young modulus
  seg = rL / N
  areaMom = bilpi * (rD / 2) ^ 4 / 4
  Cells(1, 5) = "error"
  Cells(1, 6) = "Y"
  i = 0
  Do
    i = i + 2
```

```
   Cells(i, 5) = " "
   Cells(i, 6) = " "
Loop Until i = 1000     '
i = 0
Do
   i = i + 1
   If i = 1 Then
      konst = seg ^ 3 * rLd * g / (areaMom * Yinit)
      j = 1
      Do
         j = j + 1
         jum = 0
         k = 0
         Do
            k = k + 1
            jum = jum + k * Cos(teta(k))
         Loop Until k = j - 1
         teta(j) = teta(j - 1) - konst * jum
      Loop Until j = N
      beda = teta(N) - tetaF
      If beda >= 0 Then
         Ybottom = 0
         Ytop = Yinit
      Else
         Ybootm = Yinit
         Ytop = 5 * Yinit
      End If
   Else
```

```
            Ynew = (Ybottom + Ytop) / 2
            konst = seg ^ 3 * rLd * g / (areaMom * Ynew)
            j = 1
            Do
               j = j + 1
               jum = 0
               k = 0
               Do
                  k = k + 1
                  jum = jum + k * Cos(teta(k))
               Loop Until k = j - 1
               teta(j) = teta(j - 1) - konst * jum
            Loop Until j = N
            beda = teta(N) - tetaF
            If beda >= 0 Then
               Ytop = Ynew
            Else
               Ybottom = Ynew
            End If
         End If
         If i > 1 Then
            Cells(i + 1, 5) = beda
            Cells(i + 1, 6) = Ynew
         End If
      Loop Until Abs(beda) <= salah
End Sub
```

--------------

Excel Window

| | A | B | C | D | E | F | G |
|---|---|---|---|---|---|---|---|
| 1 | **Input Parameters:** | | | | error | Y | |
| 2 | | | | | | | |
| 3 | Input rod diameter (m) | 1.30E-03 | | | 0.43913 | 2.5E+09 | |
| 4 | Input rod length density (kg/m) | 0.002308 | | | 0.429444 | 1.25E+09 | |
| 5 | Input gravitation acceleration (m/s^2) | 9.77 | | | 0.410019 | 6.25E+08 | |
| 6 | Input the number of segments | 100 | | | 0.370964 | 3.13E+08 | |
| 7 | Input rod length (m) | 0.1 | | | 0.292089 | 1.56E+08 | |
| 8 | Input angle of the free end (rad) | 0.448799 | | | 0.131682 | 78125000 | |
| 9 | Input the fixed angle (rad) | 0 | | | -0.19617 | 39062500 | |
| 10 | Input angle error (rad) | 0.0001 | | | 0.023178 | 58593750 | |
| 11 | Input initial estimation of the Young's Modulus (Pa) | 5.00E+09 | | | -0.06428 | 48828125 | |
| 12 | | | | | -0.01652 | 53710938 | |
| 13 | Calculated area moment (m^4) | 1.40E-13 | | | 0.004208 | 56152344 | |
| 14 | | | | | -0.00592 | 54931641 | |
| 15 | | | | | -0.0008 | 55541992 | |
| 16 | | | | | 0.001718 | 55847168 | |
| 17 | | | | | 0.000463 | 55694580 | |
| 18 | | | | | -0.00017 | 55618286 | |
| 19 | | | | | 0.000148 | 55656433 | |
| 20 | | | | | -9.7E-06 | 55637360 | |
| 21 | | | | | | | |
| 22 | | | | | | | |

## Supplementary 2

Mesurement the change of spaghetti rod lengths as function of time. We used two rods hung vertically.

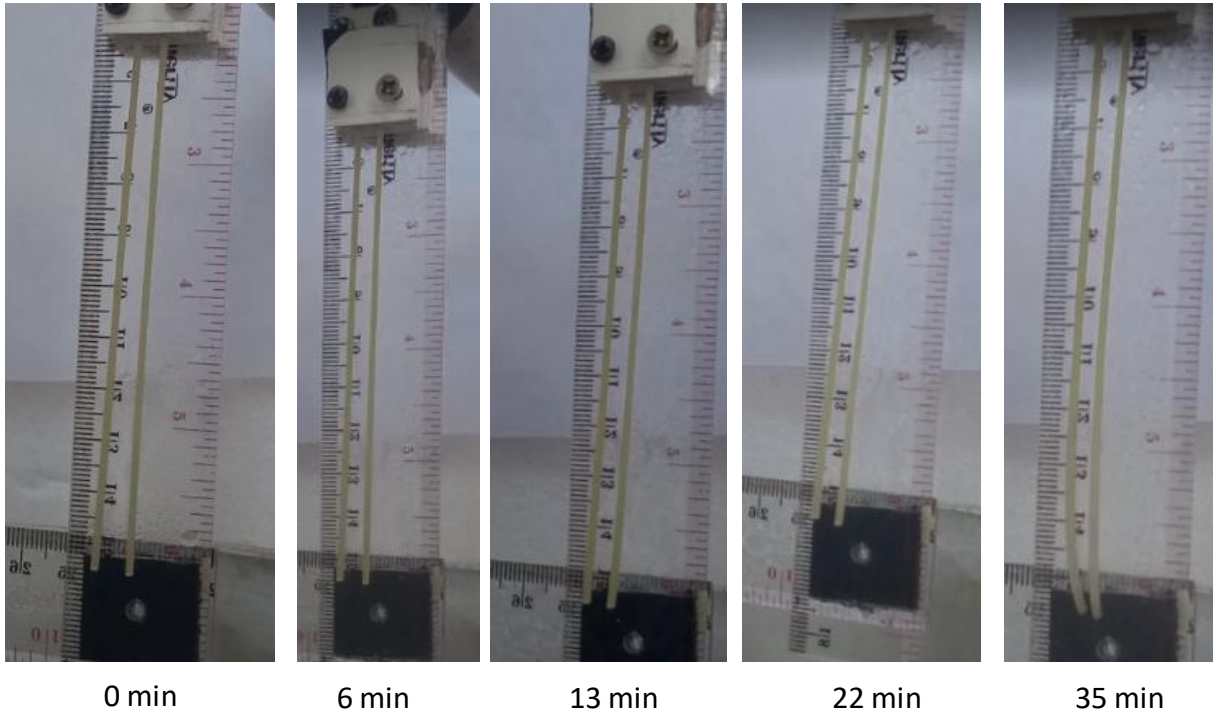

| 0 min | 6 min | 13 min | 22 min | 35 min |

The result of fraction of rod lengths at different time. Sample exposed for 35 min shows curly botom end so the the length (distance from ened to end was slammer that samle of 22 min)

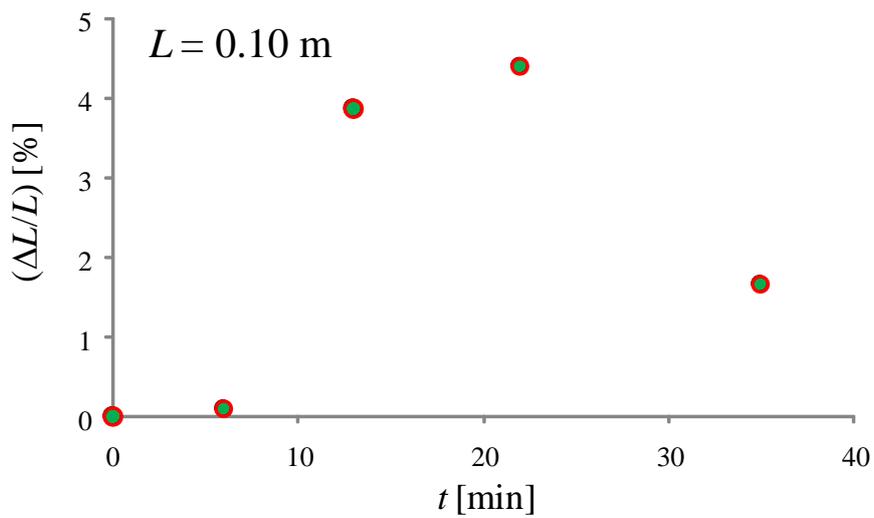

Suplementary 3

Comparison of the observed rod bending and the simulated curves. The Young's modulus at each figure is that used in calculation.

$t = 7.68$ min
$Y = 0.236$ GPa

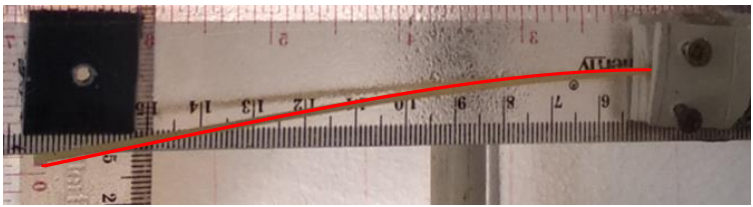

$t = 10.07$ min
$Y = 0.165$ GPa

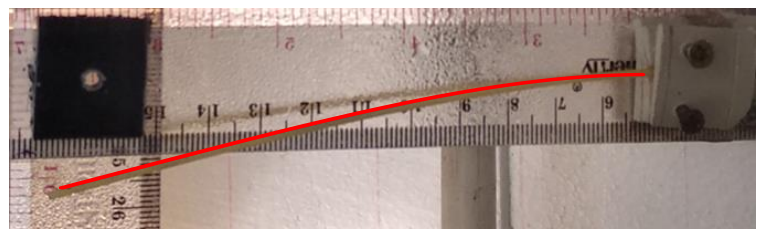

$t = 12.72$ min
$Y = 0.114$ GPa

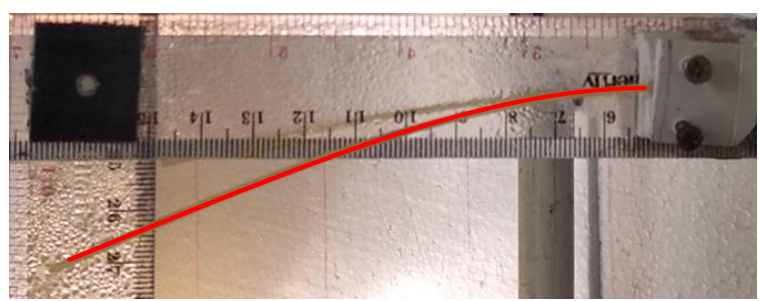

$L = 0.12$ m

$t = 16.38$ min
$Y = 0.085$ GPa

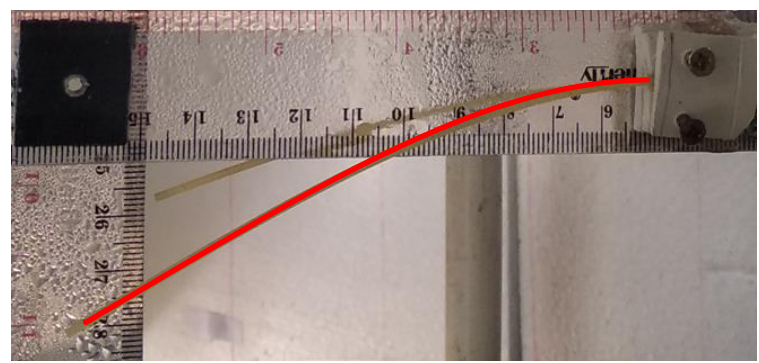

## Supplementary 4

Measured deflections of the free end of the dry spaghetti rod as function of load. This observation was used to estimate the Young's modulus of the dry spaghetti. We fit the data with a straight line and obtained a fitting curve of y = 1.425 x. From this slope we estimated the Young's modulus based on the slope = $L^3/3YI$.

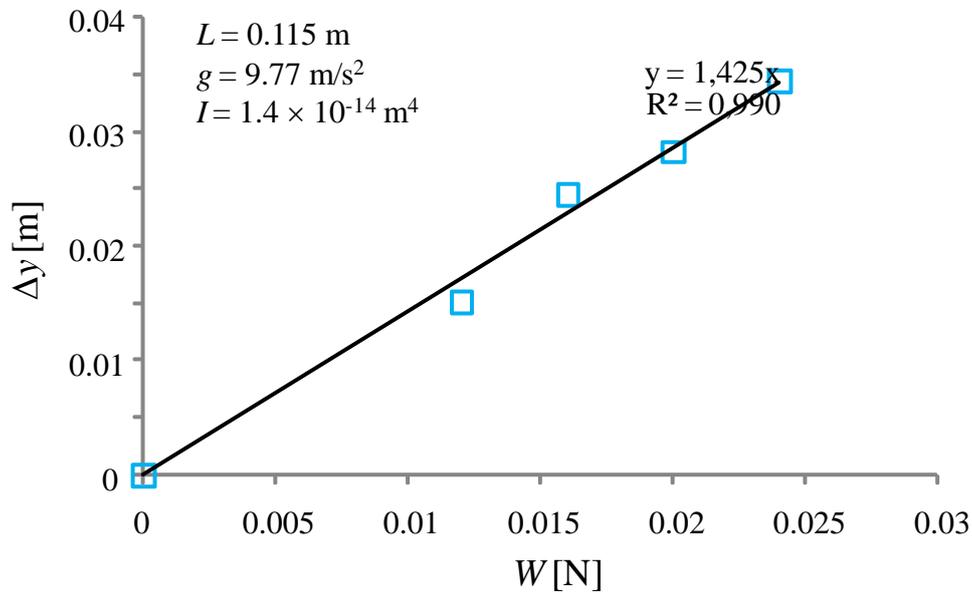

## Supplementary 5

Green curve is the upperbound curve and red curve is the lowerbound curve for fitting the bending beam. Green curve has been simulated using $Y = 41.5$ MPa and the red curve has been simulated using $Y = 39.5$ MPa. The spaghetti beam has a length of 0.10 m. From this data we concluded the true Young's modulus is located between 39.5 MPa and 41.5 MPa. This means that the error in estimating the Young's modulus is around (41.5 – 39.5)×100%/40.5 = 5%.

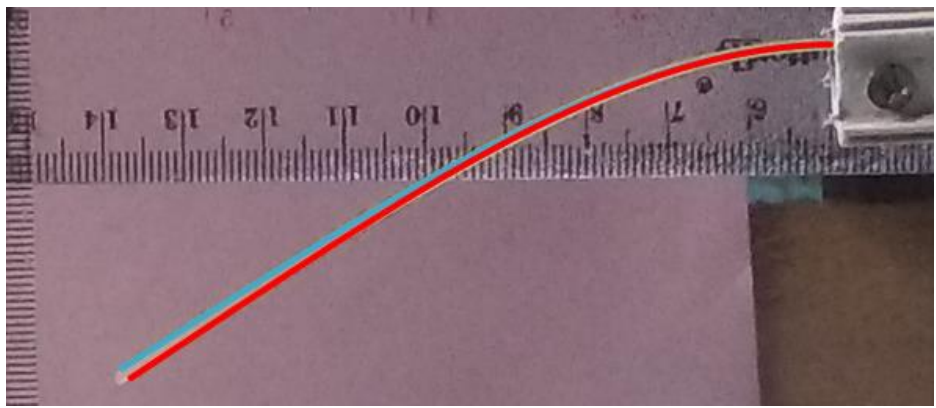